\input{aipcheck}

\documentclass[
    ,final            
  ]
  {aipproc}

\layoutstyle{6x9}

\newcommand{\bp}{{\bf{p}}}

\newcommand{\br}{{\bf{r}}}

\newcommand{\bkappa}{{\bf{\kappa}}}
\newcommand{\bb}{{\bf{b}}}

\begin{document}

\title{Multipomeron Cuts and Hard Processes on Nuclei}

\classification{12.38.Bx, 11.80.La, 13.85.Hd}
\keywords      {QCD,Multiple scattering,Diffraction}

\author{Wolfgang Sch\"afer}{
  address={Institute of Nuclear Physics PAN, ul. Radzikowskiego 152, 
31-342 Krak\'ow, Poland}
}

\begin{abstract}
With nuclear targets comes a new scale into the pQCD description of hard processes
-- the saturation scale. In the saturation regime, the familiar linear 
$k_\perp$--factorization breaks down and must be replaced by a nonlinear 
$k_\perp$--factorization, which accounts for absorptive and multiple scattering 
corrections to the hard process. Predictions for partial cross sections 
corresponding to a fixed number of cut Pomerons (the topological cross sections)
can be obtained in a surprisingly straightforward manner.
We discuss some applications to deep inelastic scattering.
\end{abstract}

\maketitle


\section{Linear $k_\perp$ factorization is broken}

Heavy nuclei are strongly absorbing targets and bring a new scale into
the perturbative QCD (pQCD) description of hard processes \cite{Mueller}.
This has severe consequences for the relations between various 
hard scattering observables. In a regime of small absorption,
small--$x$ processes are adequatly described by the 
linear $k_\perp$--factorization, and the pertinent observables
are linear functionals of a universal unintegrated gluon distribution.
Not so for the strongly absorbing target, where the linear 
$k_\perp$--factorization is broken \cite{DIS_Dijets}, 
and has to be replaced by a new, nonlinear 
$k_\perp$--factorization \cite{DIS_Dijets,Nonuniversality}, 
where observables are in general
nonlinear functionals of a properly defined unintegrated glue.
The relevant formalism has been worked out for all interesting 
processes \cite{Nonuniversality,Nonlinear} (see \cite{CGC} 
for references on related work), but in this 
very short contribution we concentrate on deep
inelastic scattering (DIS). Here, in the typical inelastic DIS event 
the nuclear debris will be left in a state with 
multiple color excited nucleons after the $q\bar q$ dipole exchanged
many gluons with the target. The partial cross sections for final states 
with a fixed number of color excited nucleons are the
topological cross sections. It is customary to describe 
them in a language of unitarity cuts through
multipomeron exchange diagrams \cite{AGK}.  
In our approach \cite{Cutting_Rules}, color excited 
nucleons in the final state give a clear--cut definition of a cut pomeron.
Topological cross sections carry useful information on the correlation between 
forward or midrapidity
jet/dijet production and multiproduction in the nuclear fragmentation region
as well as on the centrality of a collision.

\section{Nuclear collective glue and its unitarity cut interpretation}

The basic ingredient of the nonlinear $k_\perp$--factorization is
the collective nuclear unintegrated glue, which made its first appearance
in our work on the diffractive breakup of pions into jets $\pi A \to 
\mathrm{jet}_1 \mathrm{jet}_2 A$ \cite{NSS}.
Indeed, in the high energy limit, the nearly back--to--back
jets acquire their large transverse momenta directly from gluons.
It is then natural use the diffractive $S$--matrix of a $q \bar q$--dipole
$S_A(\bb,x,\br)$ for defining the nuclear unintegrated glue:
\begin{eqnarray}
\int {d^2 \br \over (2 \pi)^2} \, 
S_A(\bb,x,\br) \exp(-i\bp \br) =
S_A(\bb,x,\br \to \infty) \delta^{(2)}(\bp) + \phi(\bb,x, \bp)
\equiv \Phi(\bb,x,\bp) \, .
\end{eqnarray}
Notice that it resums multiple scatterings of a dipole, so that there 
is no straightforward relation to the conventional parton distribution
which corresponds to just two partons in the $t$--channel.
It is still meaningful to call it an unintegrated glue 
-- one reason was given above -- another one, besides its role in 
factorization formulas is its small-$x$ evolution property:
The so--defined $\phi(\bb,x,\bp)$ can be shown \cite{NS_LPM} 
to obey \footnote{Strictly speaking
only a few iterations of this equation make good sense.} 
the Balitskii--Kovchegov \cite{BK} evolution equation. 

Close to $x_A \sim (m_N R_A)^{-1}$, for heavy nuclei, the dipole $S$--matrix is the familiar Glauber--Gribov exponential
$S_A(\bb,x_A,\br) = \exp[-\sigma(x_A,\br)T(\bb)/2]$; for large dipole sizes it 
can be expressed as $S_A(\bb,x_A,\br \to \infty) = \exp[-\nu_A(x_A,\bb)]$.
Here the nuclear opacity 
$\nu_A(x_A,\bb) = {1 \over 2 } \sigma_0(x_A) T(\bb)$, is
given in terms of the dipole cross section for large dipoles 
$\sigma_0(x) = \sigma(x_A,\br \to \infty)$.
In momentum space, a useful expansion is in terms of multiple convolutions
of the free--nucleon unintegrated glue
(we use a notation 
$f(x,\bp) \propto \bp^{-4} \partial G(x,\bp^2)/ \partial \log(\bp^2)$):
\begin{eqnarray}
\phi(\bb,x_A,\bp) = \sum_{j \geq 1}  w_j \big(\nu_A(x_A,\bb)\big)
 f^{(j)}(x_A,\bp) \, . 
\label{Nuc_glue}
\end{eqnarray}
Here 
\begin{equation}
w_j(x_A,\nu_A) = {\nu_A^j(x_A,\bb) \over j! } \exp[-\nu_A(x_A,\bb)]
, \,
f^{(j)} (x_A,\bp) = \displaystyle \int \big[ \prod^j d^2 \bkappa_i 
f(x_A,\bkappa_i) \big] \delta^{(2)}( \bp - \sum \kappa_i)   \, . 
\end{equation}

Curiously, the very same collective nuclear glue is proportional to the
spectrum of quasielastically scattered quarks:
\begin{equation}
{d\sigma(qA\to qX) \over d^2 \bb d^2 \bp} 
\propto \phi(\bb,x_A,\bp) \, .
\end{equation}

Now, we can state the {\bf{first unitarity cutting rule}} in momentum space:
the $k$--th order term in the expansion (\ref{Nuc_glue}) 
corresponds to the topological cross section for the quark--nucleus 
scattering with $k$ color excited nucleons in the final state:
\begin{equation}
{d \sigma^{(k)} (qA \to qX) \over d^2 \bb d^2 \bp} \propto 
w_k \big(\nu_A(\bb) \big) f^{(k)} (\bp)  \, .
\end{equation}
This simple substitution rule forms at the heart of the cutting
rules applied to the nonlinear quadratures of \cite{Nonlinear}.

\section{Standard AGK vs. QCD}

Given the close relation between the nuclear unintegrated glue and
the Glauber--Gribov scattering theory from color dipoles, one may
be tempted to play around with various expansions of the exponential. 
Taking
inspiration from 1970's hadronic models one may then 'derive' 
expressions for topological cross sections.
For example, the inelastic cross section of the 
$q \bar{q}$-dipole-nucleus interaction is certainly obtained from:
\begin{eqnarray}
\Gamma^{inel} (\bb, \br) &=& 1 - \exp[- \sigma(\br) T(\bb)] 
= \sum_k \Gamma^{(k)}(\bb,\br) \, ,
\end{eqnarray}
and $\Gamma^{(k)}(\bb,\br) = \exp[-\sigma(\br) T(\bb)]
(\sigma(\br) T(\bb))^k/k!$ is then interpreted as the $k$--cut Pomeron
topological cross section. This is entirely incorrect, 
the reason is that this result neglects the color--coupled channel
structure of the intranuclear evolution of the color dipole.
Interestingly, a simple closed expression can be obtained
with full account for color \cite{Cutting_Rules}:
\begin{eqnarray}
\Gamma^{(k)}(\bb,\br) = \sigma(\br) T(\bb) \, w_{k-1}(2 \nu_A(\bb) )
 {e^{-2 \nu_A(\bb)} \over \lambda^k} \gamma(k,\lambda) 
\, ,
\nonumber
\end{eqnarray}
where $ \lambda = 2 \nu_A(\bb) - \sigma(\br) T(\bb)$, 
and $\gamma(k,x)$ is an incomplete Gamma--function.
For a more quantitative comparison, consult fig 1. We see that
the standard Glauber--AGK predicts a strong hierarchy: $k$ cuts 
are suppressed by the $k$--th power of the dipole cross section.
In the QCD--cutting rules there is an additional dimensionful parameter,
the opacity of a nucleus for large dipoles $\nu_A$, 
and the distribution over $k$ is substantially broader.
This difference will be more dramatic the smaller the dipole and
reflects itself in the predicted $Q^2$--dependence of 
DIS structure functions with fixed multiplicity of cut Pomerons. More
figures, as well as another example for the failure of standard AGK,
can be found on the conference website.

\section{Conclusions}
Topological cross sections can be obtained from nonlinear $k_\perp$ 
factorization formulas by straightforward substitution (cutting) rules.
For a correct isolation of topological cross sections a careful
treatment of the color coupled channel properties of the color(!) dipole
intranuclear evolution is mandatory. Don't be misguided by 
simple formulas derived in a single channel context, or by a too
literal analogy between color transparency and the Chudakov--Perkins
suppression of multiple ionisation by small size $e^+ e^-$ pairs in QED.
 

\begin{figure}
  \includegraphics[height=.4\textheight, angle=270]{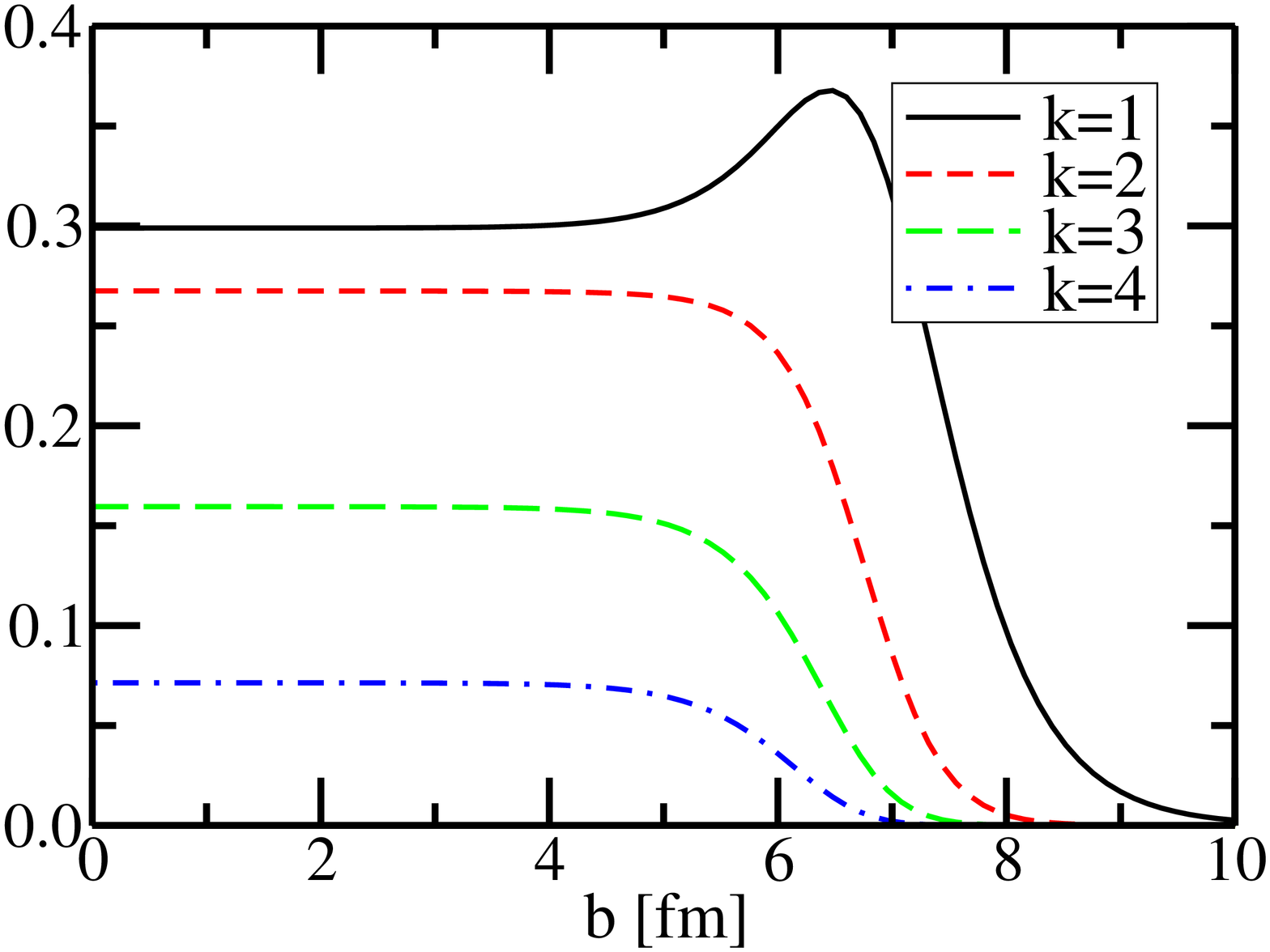}
  \includegraphics[height=.4\textheight, angle=270]{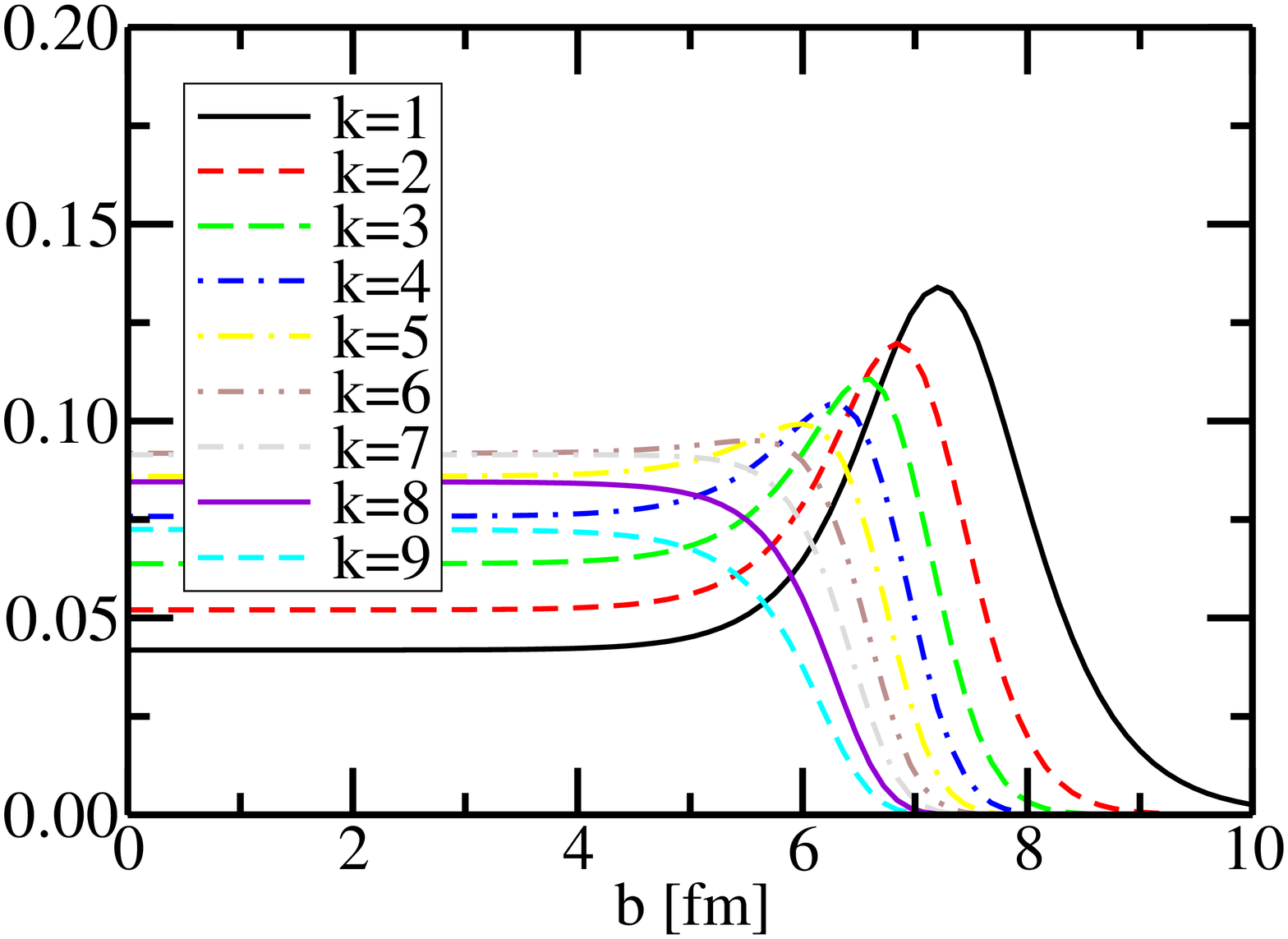}
  \caption{{\bf{Left}}: the profile function for $k$ cut Pomerons according to standard Glauber--AGK for a fairly large dipole $r = 0.6$ fm at $x=0.01$ for $A=208$. {\bf{Right}}: the same for the QCD cutting rules.}
\end{figure}

\begin{theacknowledgments}
 It is a pleasure to thank the organizers for the kind invitation.
This work was partially supported by the Polish Ministry for Science
and Higher Education (MNiSW) under contract 1916/B/H03/2008/34.
\end{theacknowledgments}



\bibliographystyle{aipproc}   



\end{document}